\documentclass[aps,prb,reprint,twocolumn,superscriptaddress,amsmath,longbibliography,nofootinbib]{revtex4-2}
\usepackage{amsmath}
\usepackage{amsfonts}
\usepackage{amssymb}
\usepackage{braket}
\usepackage{tcolorbox}
\usepackage{xcolor}
\usepackage{graphicx}
\usepackage{enumerate}

\begin{document}

\title{Sub-ballistic operator growth in spin chains with heavy-tailed random fields}
\author{Christopher L. Baldwin} \email{baldw292@msu.edu}
\affiliation{Department of Physics and Astronomy, Michigan State University, East Lansing, Michigan 48824, USA}
\date{\today}

\begin{abstract}
We rigorously prove that in nearly arbitrary quantum spin chains with power-law-distributed random fields, namely such that the probability of a field exceeding $h$ scales as $h^{-\alpha}$, it is impossible for any operator evolving in the Heisenberg picture to spread with dynamical exponent less than $1/\alpha$.
In particular, ballistic growth is impossible for $\alpha < 1$, diffusive growth is impossible for $\alpha < 1/2$, and any finite dynamical exponent becomes impossible for sufficiently small $\alpha$.
This result thus establishes a wide family of models in which the disorder provably prevents conventional transport.
We express the result as a tightening of Lieb-Robinson bounds due to random fields --- the proof modifies the standard derivation such that strong fields appear as effective weak interactions, and then makes use of analogous recent results for random-bond spin chains.
\end{abstract}

\maketitle

\section{Introduction} \label{sec:introduction}

A large amount of work in recent years has been devoted to whether and how quenched disorder inhibits the ``natural'' dynamics of isolated quantum many-body systems.
Of particular interest is the phenomenon of many-body localization (MBL), in which strong disorder (say due to random fields in a spin chain) gives rise to quasi-local conserved quantities that prevent the system from coming to equilibrium~\cite{Nandkishore2015Many,Alet2018ManyBody,Abanin2019Colloquium,Sierant2024ManyBody}.
MBL has a rich phenomenology that has been extensively developed over the past two decades~\cite{Oganesyan2007Localization,Znidaric2008Many,Pal2010Many,Bardarson2012Unbounded,Vosk2013ManyBody,Huse2013Localization,Serbyn2013Local,Pekker2014HilbertGlass,Nandkishore2014Spectral,Huse2014Phenomenology,Ros2015Integrals,Chandran2015Constructing,Agarwal2015Anomalous,Gopalakrishnan2015Low,Vosk2015Theory,DeRoeck2016Absence,Chandran2016Many,DeRoeck2017Stability,Dumitrescu2019KosterlitzThouless,Morningstar2019RenormalizationGroup,Crowley2020Avalanche,Garratt2021ManyBody,Garratt2021Local,Crowley2022MeanField} (building on a history of earlier works~\cite{Lee1985Disordered,Altshuler1997Quasiparticle,Gornyi2005Interacting,Basko2006Metal}).

Most work has focused on MBL in one dimension, as it is argued to be unstable (at least in the strictest sense) in higher dimensions~\cite{DeRoeck2017Stability}.
Yet even in one dimension, the existence of MBL in physically motivated models remains controversial, as one can appreciate from the recent review on the subject~\cite{Sierant2024ManyBody}.
The majority of research has been numerical, limited to either small system sizes or finite evolution times (as is also true of experimental studies~\cite{Schreiber2015Observation,Kondov2015Disorder,Smith2016Many,Wei2018Exploring}), and thus is subject to multiple competing interpretations~\cite{Suntajs2020Quantum,Abanin2021Distinguishing,Sels2021Dynamical,Sierant2022Challenges,Sels2023Thermalization}.
Perturbative calculations rely on uncontrolled approximations~\cite{Basko2006Metal,Ros2015Integrals}, and even a proposed mathematical proof~\cite{Imbrie2016On} has not settled the debate (see also Ref.~\cite{DeRoeck2023Rigorous}).
Recent works, therefore, tend to be cautious in referring to MBL ``regimes'' (rather than phases), so as to highlight that the fate at large sizes and/or times remains unclear~\cite{Morningstar2022Avalanches,Crowley2022Constructive,Long2023Phenomenology}.

The present paper does \textit{not} resolve the controversy surrounding MBL --- it instead identifies an analogous yet simpler situation in which strong disorder can be rigorously shown to suppress the natural dynamics of one-dimensional spin chains.
Specifically, we consider spin chains with \textit{power-law}-distributed random fields.
We prove that when the fields have a sufficiently heavy tail, ballistic and even diffusive motion (in an extremely general sense) becomes impossible.
This is true regardless of the observables in question, the types of interactions, and even any driving terms.

\begin{figure}[t]
\centering
\includegraphics[width=0.85\columnwidth]{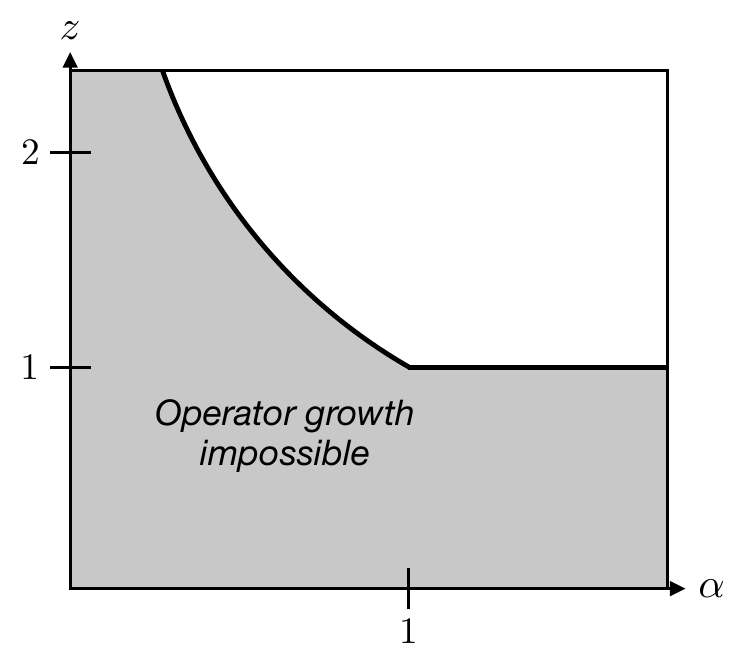}
\caption{Sketch of the main result: when a spin chain has random fields distributed as a power law with exponent $\alpha$, then it is impossible for any operator to grow with a dynamical exponent $z$ in the gray region, namely $z < \max \{1/\alpha, 1\}$. The x-axis is $\alpha$: the probability of a field exceeding $h$ scales as $h^{-\alpha}$ at large $h$. The y-axis is $z$: we consider whether an operator can spread across a distance $r$ in a time $t \propto r^z$, finding that it cannot for $z$ in the gray region. Note that this does not necessarily imply that operator growth \textit{is} possible throughout the white region.}
\label{fig:phase_diagram}
\end{figure}

To be more precise, we study the time evolution of initially local operators $A_x$ (say acting on site $x$ of the chain) and identify lower bounds on the time needed for $A_x(t)$ to act appreciably on a different site $y$.
Suppose the random fields in the chain are such that the probability of a field exceeding $h$ scales as $h^{-\alpha}$ at large $h$.
We prove that there is a minimum possible dynamical exponent $z_c = \max \{1/\alpha, 1\}$ for the growth of $A_x(t)$, in that it is impossible for $A_x(t)$ to act significantly on $y$ in a time of order $|y-x|^z$ for any $z < z_c$.
This is sketched in Fig.~\ref{fig:phase_diagram}, with the gray region indicating those dynamical exponents that are impossible to attain.

This result is naturally expressed in terms of Lieb-Robinson (LR) bounds, which have become their own topic of study in recent years.
Following the original discovery~\cite{Lieb1972Finite}, they were used (much later) to rigorously prove some important folk theorems in quantum many-body physics~\cite{Hastings2004LiebSchultzMattis,Nachtergaele2006LiebRobinson,Hastings2006Spectral}, and have since found an array of applications in quantum information~\cite{Bravyi2006LiebRobinson,Osborne2006Efficient,Eldredge2017Fast,Haah2018Quantum,Huang2022Provably}.
See Ref.~\cite{Chen2023Speed} for a recent review.
Accordingly, LR bounds have been both generalized and specialized in numerous ways over recent years: for example, tightening the bounds due to certain interaction structures~\cite{Wang2020Tightening}, expanding to long-range interactions~\cite{FossFeig2015Nearly,Tran2020Hierarchy,Tran2021LiebRobinson}, extending to open systems~\cite{Poulin2010LiebRobinson,Barthel2012Quasilocality}, and combinations thereof~\cite{Sweke2019LiebRobinson,Guo2021Clustering,Roon2024QuasiLocality}.
The present work fits neatly into this line of studies, showing that LR bounds can be dramatically tightened by accounting for random fields.

This work also complements a number of recent results concerning slow dynamics in random-\textit{bond} models, motivated by a search for non-ergodic phenomena in regimes where conventional MBL does not apply (say due to non-abelian symmetries or in a classical limit).
Refs.~\cite{Protopopov2020NonAbelian,McRoberts2023Subdiffusive,Saraidaris2024FiniteSize,Dabholkar2024Ergodic} specifically consider power-law-distributed interaction strengths, and find that anomalous behaviors such as sub-diffusion and slow thermalization result when the distribution has a sufficiently heavy tail.
Also relevant are works studying dynamics in disordered kinetically-constrained models~\cite{Chen2018How}, where random ``fields'' can play a role similar to random interactions due to the constraints.
Along these lines, Ref.~\cite{Baldwin2023Disordered} derives improved LR bounds for random-bond spin chains, proving that the shape of the ``light cone'' is modified when --- much as in Refs.~\cite{Protopopov2020NonAbelian,McRoberts2023Subdiffusive,Saraidaris2024FiniteSize,Dabholkar2024Ergodic} --- the distribution of interaction strengths is heavy-tailed.
The present work builds naturally on the techniques and results found in Ref.~\cite{Baldwin2023Disordered}.

The conventional LR bound proves that any initially local observable, as it evolves in time (in the Heisenberg picture) and becomes non-local, cannot spread faster than ballistically.
In other words, if an operator $A_x$ initially acts on site $x$, then $A_x(t)$ cannot act appreciably on site $y$ in a time less than $|y-x|/v$ for some velocity $v$ --- the LR bound further identifies an upper bound $v_{\textrm{LR}}$ on the allowed velocity.
The upper bound depends only on the lattice geometry and an overall interaction strength --- in the conventional setting, all interactions are taken to be of order $J$, yielding $v_{\textrm{LR}}$ that is of order $J$.
This existing result already accounts for the region $\alpha \geq 1$ in Fig.~\ref{fig:phase_diagram}, and accordingly we will focus specifically on the case $\alpha < 1$ in this paper.

Ref.~\cite{Baldwin2023Disordered} in turn considers one-dimensional spin chains in which the interaction strength $J_{j,j+1}$ between sites $j$ and $j+1$ is distributed as a power-law --- the probability of $J_{j,j+1} < J$ scales as $J^{\alpha}$ at small $J$ --- and proves that $A_x(t)$ cannot act appreciably on $y$ in a time less than $O(|y-x|^{1/\alpha})$ when $\alpha < 1$.
Thus ballistic operator growth is impossible for all $\alpha < 1$, diffusive growth is impossible for $\alpha < 1/2$, and so on.
The result has a straightforward explanation, at least qualitatively: the weakest interaction within distance $r \equiv |y-x|$, denoted $J_{\textrm{min}}(r)$, is of order $r^{-1/\alpha}$, and thus at a minimum, it takes time of order $J_{\textrm{min}}(r)^{-1} \sim r^{1/\alpha}$ for an operator to spread across that link of the chain.
This is thus an example of a rare-region effect, one that is exceedingly general since it is agnostic to all other details of the Hamiltonian.
The challenge is in proving that there is no way to circumvent this bottleneck, especially since prior LR bounds (and closely related calculations such as short-time expansions) suggest that there should be.
See Ref.~\cite{Baldwin2023Disordered} for further details.

Here we return to spin chains that do not necessarily have weak interactions, but rather have power-law random fields $h_j$ --- the probability of $|h_j| > h$ scales as $h^{-\alpha}$ at large $h$.
We prove that here as well, $A_x(t)$ cannot act appreciably on $y$ in a time less than $O(|y-x|^{1/\alpha})$ when $\alpha < 1$.
This follows naturally from Ref.~\cite{Baldwin2023Disordered}: it is well-known that strong fields of order $h$ can function as effective weak interactions of order $J^2/h$, as formalized in frameworks such as the strong-disorder renormalization group~\cite{Fisher1995Critical,Refael2013Strong}, and so power-law-distributed fields in particular amount to effective interactions of the type for which Ref.~\cite{Baldwin2023Disordered} applies.
Once again, the challenge is in making this argument precise and sufficiently general (not least because local fields do not enter into the conventional LR bounds to begin with).

The question of whether strong local fields tighten LR bounds has in fact recently been considered in the mathematical literature as well~\cite{Gebert2022LiebRobinson}.
While the language and notation are rather different, certain elements of the proof in Ref.~\cite{Gebert2022LiebRobinson} are strikingly similar to ours, in particular the technique for converting strong fields to weak interactions.
Yet differences in the subsequent analysis mean that Ref.~\cite{Gebert2022LiebRobinson}, while deriving a substantially smaller prefactor for the LR bound, does not identify a change in the LR velocity, whereas here we prove that even the ballistic nature of the LR bound is in fact modified.
It would be very interesting to consider whether the two approaches can be merged to obtain stronger results.

Even more recently, Ref.~\cite{DeRoeck2024Absence} has supplied a proof that conductivities vanish in a wide family of strongly disordered spin chains.
The results are both analogous and complementary to those given here.
On the one hand, Ref.~\cite{DeRoeck2024Absence} studies models with much milder field distributions yet in the limit where interactions are weaker even than typical fields (this is the situation most relevant for MBL).
Here, by contrast, we consider field distributions with very heavy tails.
In particular, our results continue to hold when interactions are \textit{stronger} than the typical fields --- it is only the rare fields in the tail of the distribution that are necessary.
With those differences in mind, it is interesting to note that the end results are rather similar: proving the impossibility of certain dynamical exponents in such a way that any finite exponent is ruled out at sufficiently weak interactions/heavy tails.
It would again be worthwhile to investigate whether these two approaches can be combined, particularly whether the analysis of Ref.~\cite{DeRoeck2024Absence} can be extended to the situations considered here.

Lastly, there are a number of other mathematical works at the intersection of LR bounds and disordered spin chains which are not as directly connected to the present study but are worth highlighting: Ref.~\cite{Burrell2009Information} considers random fields which fluctuate rapidly in time, Refs.~\cite{Burrell2007Bounds,Hamza2012Dynamical,Gebert2016On,Elgart2018Manifestations} construct LR bounds for specific (often free-fermion-integrable) models, Refs.~\cite{Kim2014Local,Nachtergaele2021Slow,Toniolo2024Stability,Toniolo2024Dynamics} establish relationships between zero-velocity LR bounds and the phenomenology of MBL, and Ref.~\cite{DeRoeck2020Subdiffusion} proves sub-diffusion in spin chains with single-particle-localized segments.

The proof of our result is in Sec.~\ref{sec:proof}.
First, however, we give a precise statement of the problem and establish notation in Sec.~\ref{sec:preliminaries}, as well as provide a short derivation of the conventional LR bound (since it will be needed for our proof).
We conclude in Sec.~\ref{sec:conclusion} by discussing some open questions.

\section{Preliminaries} \label{sec:preliminaries}

\subsection{Setup} \label{subsec:setup}

Our main result applies to one-dimensional spin chains in a very broad sense: the spins can be arbitrary $d$-state degrees of freedom (``qudits''), the interactions can take nearly any form (even time-dependent), and the fields can couple to any local operators as long as they are non-degenerate.
In precise terms, we are considering any $N$-site system with Hamiltonian $H(t)$ of the form
\begin{equation} \label{eq:general_Hamiltonian_form}
H(t) = \sum_{j=1}^{N-1} H_{j,j+1}(t) - \sum_{j=1}^N h_j \hat{s}_j,
\end{equation}
subject to the following conditions:
\begin{itemize}
\item The interaction term $H_{j,j+1}(t)$, which acts only on spins $j$ and $j+1$, has bounded magnitude and time-derivative~\cite{time_derivative_note}:
\begin{equation} \label{eq:interaction_norm_bounds}
\big\lVert H_{j,j+1}(t) \big\rVert \leq J, \qquad \big\lVert \partial_t H_{j,j+1}(t) \big\rVert \leq \zeta J,
\end{equation}
where $\lVert \, \cdot \, \rVert$ denotes the operator norm (largest eigenvalue in absolute value).
\item The operator $\hat{s}_j$, which acts only on spin $j$, is non-degenerate and time-independent.
It also has bounded magnitude:
\begin{equation} \label{eq:local_norm_bound}
\big\lVert \hat{s}_j \big\rVert \leq b.
\end{equation}
\item The field $h_j$ is random, drawn independently of all others from a distribution such that, at large $h$~\cite{notation_note},
\begin{equation} \label{eq:local_field_distribution}
\textrm{Pr} \big[ |h_j| > h \big] \sim Ch^{-\alpha},
\end{equation}
for some constant $C$ and $\alpha > 0$ (we use the notation $\textrm{Pr}[ \, \cdot \, ]$ to denote the probability of the event in question).
Note that the power-law distribution in Eq.~\eqref{eq:local_field_distribution} is somewhat different than that usually considered in studies of MBL (which rather tends to be uniform over a scale $W \gg J$).
\end{itemize}

We prove that any initially local operator $A_x$, evolving in the Heisenberg picture under the Hamiltonian in Eq.~\eqref{eq:general_Hamiltonian_form}, cannot ``spread out'' with a dynamical exponent less than $1/\alpha$ when $\alpha < 1$.
To state this result precisely, it is useful to establish some convenient notation within the ``superoperator'' formalism.

\subsection{Definitions and notation} \label{subsec:definitions}

Many recent works on dynamics in spin-1/2 chains have found it convenient to expand observables as linear combinations of ``strings'' of Pauli operators (e.g., $\hat{\sigma}_1^x \otimes \hat{\sigma}_2^y \otimes I_3 \otimes \hat{\sigma}_4^z \otimes \cdots$) and study how the coefficients of the strings evolve in time.
This can be generalized to any model involving arbitrary degrees of freedom (we will refer to all such models as ``spin chains'' from now on), and it gives a picture of the operator dynamics that is somewhat analogous to the dynamics of a \textit{single} particle on the lattice:
\begin{itemize}
\item \textbf{Basis states:} For a single particle hopping on a lattice, the sites of the lattice define a natural basis for the wavefunction of the particle.
For a spin chain, we can similarly construct an orthonormal basis for observables.
Hermitian operators form a real vector space with the inner product between $A$ and $B$ given by the trace product, $d^{-N} \textrm{Tr} AB$.
Thus first consider a single site $j$ of the chain, and choose an orthonormal basis denoted by $X_j^{\nu_j}$ ($\nu_j \in \{0, \cdots, d^2-1\}$ with $d$ the local Hilbert space dimension).
We take $X_j^0$ to be the identity $I_j$, but otherwise the basis is arbitrary.
Then take the basis $X^{\nu}$ for the entire chain to be tensor products (``strings'') of the single-site basis operators, labeled by $\nu \equiv (\nu_1, \cdots, \nu_N)$:
\begin{equation} \label{eq:basis_operators_definition}
X^{\nu} \equiv \bigotimes_{j=1}^N X_j^{\nu_j}.
\end{equation}
Any observable $A$ can be written as a linear combination of $X^{\nu}$, i.e., $A = \sum_{\nu} c_{\nu} X^{\nu}$ with real coefficients $c_{\nu}$.
\item \textbf{Projectors:} For a single particle, and for any subset $\omega$ of sites on the lattice, the operator $\sum_{j \in \omega} | j \rangle \langle j |$ projects any state into $\omega$, such that the norm (squared) of the projected state gives the probability of the particle being observed in $\omega$.
For a spin chain, similarly define a projection superoperator $\mathcal{P}_{\omega}$ that preserves only those basis strings which have a non-identity element somewhere in $\omega$:
\begin{equation} \label{eq:projection_superoperator_definition}
\mathcal{P}_{\omega} X^{\nu} \equiv \bigg[ 1 - \prod_{j \in \omega} \delta_{\nu_j, 0} \bigg] X^{\nu}.
\end{equation}
The action of $\mathcal{P}_{\omega}$ on any operator $A$ is determined by linearity --- writing $A$ as a linear combination of basis strings, $\mathcal{P}_{\omega} A$ keeps only those terms that act non-trivially in $\omega$.
The norm of $\mathcal{P}_{\omega} A$ is thus a measure of the ``strength'' with which $A$ acts in $\omega$.
\item \textbf{Dynamics:} In the Schrodinger picture, the evolution of a quantum state $| \psi \rangle$ is governed by the Schrodinger equation, $i \partial_t | \psi(t) \rangle = H(t) | \psi(t) \rangle$.
The solution is $| \psi(t) \rangle = U(t) | \psi \rangle$ with evolution operator $U(t) \equiv \mathcal{T} \exp{[-i \int_0^t \textrm{d}s H(s)]}$, where $\mathcal{T}$ denotes time-ordering.
In the Heisenberg picture, the evolution of an observable $A$ is governed by the Heisenberg equation~\cite{Heisenberg_note}
\begin{equation} \label{eq:Heisenberg_equation}
\partial_t A(t) = i \big[ H(t), A(t) \big].
\end{equation}
Define the ``Liouvillian'' superoperator $\mathcal{L}(t)$ by $\mathcal{L}(t) A(t) \equiv i[H(t), A(t)]$, so that Eq.~\eqref{eq:Heisenberg_equation} can be written $\partial_t A(t) = \mathcal{L}(t) A(t)$.
Just as above, the solution is therefore given by the time-ordered exponential: $A(t) = \mathcal{U}(t) A$ with $\mathcal{U}(t) \equiv \mathcal{T} \exp{[\int_0^t \textrm{d}s \mathcal{L}(s)]}$.
More generally, denote by $\mathcal{U}(t, t')$ the (unitary) superoperator that evolves from time $t'$ to time $t$.
The key facts for what follows are that
\begin{equation} \label{eq:evolution_superoperator_facts}
\begin{aligned}
\partial_t \mathcal{U}(t, t') &= \mathcal{L}(t) \mathcal{U}(t, t'), \\
\partial_{t'} \mathcal{U}(t, t') &= -\mathcal{U}(t, t') \mathcal{L}(t'), \\
\partial_t \mathcal{U}(t, t')^{\dag} &= -\mathcal{U}(t, t')^{\dag} \mathcal{L}(t), \\
\partial_{t'} \mathcal{U}(t, t')^{\dag} &= \mathcal{L}(t') \mathcal{U}(t, t')^{\dag}.
\end{aligned}
\end{equation}
Note that any Hamiltonian has an associated Liouvillian and evolution superoperator --- when we use symbols to distinguish between different Hamiltonians, we will use the same symbols for the corresponding superoperators (e.g., $\overline{\mathcal{L}}$ and $\overline{\mathcal{U}}$ are the Liouvillian and evolution superoperator associated to a Hamiltonian $\overline{H}$ that will be clear from context, $\mathcal{L}_0$ and $\mathcal{U}_0$ are those associated to $H_0$, and so on).
\end{itemize}

Consider a local operator $A_x$ that acts only on site $x$.
Due to interactions, the time-evolved operator $A_x(t)$ will no longer act only on $x$, but the locality of the spin chain ensures that it will take time for $A_x(t)$ to act appreciably on distant sites.
To quantify this, we derive a bound on the projection of $A_x(t)$ past a further site $y$, i.e., a bound on $\lVert \mathcal{P}_{\geq y} A_x(t) \rVert \equiv f_x(y, t)$ (where $\geq y$ denotes those sites to the right of and including site $y$ --- we take $y > x$ without loss of generality).
The conventional result, called a ``Lieb-Robinson'' (LR) bound, is that there exists a finite velocity $v_{\textrm{LR}}$ such that
\begin{equation} \label{eq:original_Lieb_Robinson_bound}
f_x(y, t) \leq 2 \lVert A_x \rVert e^{v_{\textrm{LR}} t - |y-x|}.
\end{equation}
Thus $f_x(y, t)$ falls off rapidly outside of the ``light cone'' defined by $|y-x| = v_{\textrm{LR}} t$, and one can interpret $v_{\textrm{LR}}$ as a speed limit on the growth of the operator $A_x$.
While Eq.~\eqref{eq:original_Lieb_Robinson_bound} is only a bound and thus different calculations yield different expressions for $v_{\textrm{LR}}$, a simple analysis reproduced below gives $v_{\textrm{LR}} = 4eJ$, valid for any local $A_x$ and for any Hamiltonian such that $\lVert H_{j,j+1}(t) \rVert \leq J$.
It can be generalized to any higher-dimensional lattice as well, although with a different resulting velocity.

In this paper, we derive a substantially tighter LR bound for Hamiltonians of the form defined in and below Eq.~\eqref{eq:general_Hamiltonian_form}, making use of the random fields distributed as in Eq.~\eqref{eq:local_field_distribution}.
In particular, we identify a dynamical exponent $z_c \equiv 1/\alpha$ such that $f_x(y, t) \ll 1$ when $t = O(|y-x|^z)$ for any $z < z_c$.
Thus ballistic operator growth is impossible whenever the random fields have $\alpha < 1$, diffusive growth is impossible whenever $\alpha < 1/2$, and so on.

This result, as are most LR bounds, is sufficiently general that it immediately puts analogous constraints on physically relevant quantities such as response functions.
For example, within linear response~\cite{Chaikin1995}, the effect that a perturbation coupling to local operator $B_y$ has on a different local observable $A_x$ (acting on sites $y$ and $x$ respectively) after time $t$ is given by the response function
\begin{equation} \label{eq:linear_response_function}
\chi_{AB}(t) = i \big< \big[ A_x(t), B_y \big] \big>,
\end{equation}
with the expectation value taken in an appropriate state/ensemble.
Since $B_y$ acts only on site $y$, we can replace $A_x(t)$ with $\mathcal{P}_{\geq y} A_x(t)$ in the commutator (the basis strings annihilated by $\mathcal{P}_{\geq y}$ commute with $B_y$ anyway).
Thus
\begin{equation} \label{eq:linear_response_bound}
\begin{aligned}
\big| \chi_{AB}(t) \big| &\leq \big\lVert \big[ \mathcal{P}_{\geq y} A_x(t), B_y \big] \big\rVert \\
&\leq 2 \big\lVert \mathcal{P}_{\geq y} A_x(t) \big\rVert \big\lVert B_y \big\rVert,
\end{aligned}
\end{equation}
and a bound on $\lVert \mathcal{P}_{\geq y} A_x(t) \rVert$ immediately gives a bound on $|\chi_{AB}(t)|$.
In particular, if $A_x(t)$ cannot spread past $y$ in a time of order $|y-x|^z$, then there cannot be an appreciable response within that time either.

\subsection{Conventional LR bound} \label{subsec:conventional_LR_bound}

Since we will make use of the results in the course of our proof, here we give a simple derivation of Eq.~\eqref{eq:original_Lieb_Robinson_bound} for the Hamiltonians under consideration (given by Eqs.~\eqref{eq:general_Hamiltonian_form} through~\eqref{eq:local_field_distribution}).
The technique uses interaction-picture transformations to ``integrate out'' certain terms in the Hamiltonian, together with repeated use of the triangle inequality and submultiplicativity of the operator norm: for any operators $A$ and $B$, $\lVert A + B \rVert \leq \lVert A \rVert + \lVert B \rVert$ and $\lVert AB \rVert \leq \lVert A \rVert \lVert B \rVert$ (note in particular the corollary $\lVert [A, B] \rVert \leq 2 \lVert A \rVert \lVert B \rVert$).

To begin, define the interaction-picture transformation
\begin{equation} \label{eq:conventional_bound_field_removal_transformation}
\overline{A}_x(t) \equiv e^{i \sum_{j=1}^N h_j \hat{s}_j t} A_x(t) e^{-i \sum_{j=1}^N h_j \hat{s}_j t},
\end{equation}
so that $\overline{A}_x(t)$ obeys the equation of motion
\begin{equation} \label{eq:conventional_bound_field_removal_equation}
\partial_t \overline{A}_x(t) = i \big[ \overline{H}(t), \overline{A}_x(t) \big],
\end{equation}
where
\begin{equation} \label{eq:conventional_bound_field_removal_interactions}
\begin{aligned}
\overline{H}(t) &\equiv \sum_{j=1}^{N-1} e^{i \sum_{k=1}^N h_k \hat{s}_k t} H_{j,j+1}(t) e^{-i \sum_{k=1}^N h_k \hat{s}_k t} \\
&\equiv \sum_{j=1}^{N-1} \overline{H}_{j,j+1}(t).
\end{aligned}
\end{equation}
The fields have been eliminated at the price of transforming the interactions.
Note that since the transformation is unitary, $\overline{H}_{j,j+1}(t)$ has the same eigenvalues as $H_{j,j+1}(t)$ and thus $\lVert \overline{H}_{j,j+1}(t) \rVert = \lVert H_{j,j+1}(t) \rVert \leq J$.
It also respects the same graph structure: $\overline{H}_{j,j+1}(t)$ continues to act only on spins $j$ and $j+1$.

Next, pick any site $j$ such that $x \leq j < y$.
Separate the (transformed) Hamiltonian into the interaction between $j$ and $j+1$ on one hand and all other terms (denoted $\overline{H}_0$) on the other:
\begin{equation} \label{eq:conventional_bound_term_decomposition}
\begin{aligned}
\overline{H}(t) &= \sum_{k \neq j} \overline{H}_{k,k+1}(t) + \overline{H}_{j,j+1}(t) \\
&\equiv \overline{H}_0(t) + \overline{H}_{j,j+1}(t).
\end{aligned}
\end{equation}
Pass to another interaction picture with respect to $\overline{\mathcal{U}}_0$ (recall that $\overline{\mathcal{U}}_0$ is the evolution superoperator associated with $\overline{H}_0$) by considering the equation of motion for $\overline{\mathcal{U}}_0(t, 0)^{\dag} \overline{A}_x(t)$.
Using Eqs.~\eqref{eq:Heisenberg_equation} and~\eqref{eq:evolution_superoperator_facts}, we have
\begin{equation} \label{eq:conventional_bound_interaction_equation_differential}
\partial_t \Big( \overline{\mathcal{U}}_0(t, 0)^{\dag} \overline{A}_x(t) \Big) = \overline{\mathcal{U}}_0(t, 0)^{\dag} \overline{\mathcal{L}}_{j,j+1}(t) \overline{A}_x(t).
\end{equation}
Integrating and multiplying by $\overline{\mathcal{U}}_0(t, 0)$ gives
\begin{equation} \label{eq:conventional_bound_interaction_equation_integral}
\overline{A}_x(t) = \overline{\mathcal{U}}_0(t, 0) A_x + \int_0^t \textrm{d}t' \overline{\mathcal{U}}_0(t, t') \overline{\mathcal{L}}_{j,j+1}(t') \overline{A}_x(t').
\end{equation}
Now act on both sides with $\mathcal{P}_{\geq j+1}$.
The key observation is that since the regions $\geq j+1$ and $\leq j$ are decoupled under $\overline{H}_0$ alone, $\mathcal{P}_{\geq j+1}$ and $\overline{\mathcal{U}}_0(t, t')$ commute --- evolution under $\overline{H}_0$ cannot spread an operator from $\leq j$ into $\geq j+1$ or vice-versa (see Ref.~\cite{Baldwin2023Disordered} for a simple proof).
At the same time, $\overline{\mathcal{L}}_{j,j+1}$ (being the commutator with $\overline{H}_{j,j+1}$) annihilates any basis string that has the identity throughout the region $\geq j$, and thus $\overline{\mathcal{L}}_{j,j+1} \overline{A}_x = \overline{\mathcal{L}}_{j,j+1} \mathcal{P}_{\geq j} \overline{A}_x$.
Eq.~\eqref{eq:conventional_bound_interaction_equation_integral} thus leads to
\begin{equation} \label{eq:conventional_bound_interaction_equation_projected}
\begin{aligned}
\mathcal{P}_{\geq j+1} \overline{A}_x(t) &= \overline{\mathcal{U}}_0(t, 0) \mathcal{P}_{\geq j+1} A_x \\
&+ \int_0^t \textrm{d}t' \overline{\mathcal{U}}_0(t, t') \mathcal{P}_{\geq j+1} \overline{\mathcal{L}}_{j,j+1}(t') \mathcal{P}_{\geq j} \overline{A}_x(t').
\end{aligned}
\end{equation}
Note that the first term, $\mathcal{P}_{\geq j+1} A_x$, in fact vanishes since we chose $j$ such that $x < j+1$.
Next taking norms and using the triangle inequality --- while noting that $\overline{\mathcal{U}}_0$, being unitary, does not change the norm of any operator that it acts on --- gives
\begin{equation} \label{eq:conventional_bound_local_inequality_v1}
\big\lVert \mathcal{P}_{\geq j+1} \overline{A}_x(t) \big\rVert \leq \int_0^t \textrm{d}t' \big\lVert \mathcal{P}_{\geq j+1} \overline{\mathcal{L}}_{j,j+1}(t') \mathcal{P}_{\geq j} \overline{A}_x(t') \big\rVert.
\end{equation}
A useful technical lemma (see Ref.~\cite{Baldwin2023Disordered}) is that for any projector $\mathcal{P}_{\omega}$ and any operator $A$,
\begin{equation} \label{eq:conventional_bound_technical_lemma}
\big\lVert \mathcal{P}_{\omega} A \big\rVert \leq 2 \big\lVert A \big\rVert.
\end{equation}
Thus $\lVert \mathcal{P}_{\geq j+1} \overline{\mathcal{L}}_{j,j+1} \cdots \rVert \leq 2 \lVert \overline{\mathcal{L}}_{j,j+1} \cdots \rVert \leq 4J \lVert \cdots \rVert$ (the latter inequality because $\overline{\mathcal{L}}_{j,j+1}$ is the commutator with $\overline{H}_{j,j+1}$ and $\lVert \overline{H}_{j,j+1} \rVert \leq J$).
We are left with
\begin{equation} \label{eq:conventional_bound_local_inequality_v2}
\big\lVert \mathcal{P}_{\geq j+1} \overline{A}_x(t) \big\rVert \leq 4J \int_0^t \textrm{d}t' \big\lVert \mathcal{P}_{\geq j} \overline{A}_x(t') \big\rVert.
\end{equation}
The last step is to return from $\overline{A}_x(t)$ to $A_x(t)$.
Note that the transformation defined in Eq.~\eqref{eq:conventional_bound_field_removal_transformation}, being a product of local transformations, maps the identity operator on any site to itself.
Thus it commutes with $\mathcal{P}_{\omega}$, regardless of the subset $\omega$ of sites in question.
Thus $\lVert \mathcal{P}_{\omega} \overline{A}_x(t) \rVert = \lVert \mathcal{P}_{\omega} A_x(t) \rVert$, and we have
\begin{equation} \label{eq:conventional_bound_local_inequality_v3}
f_x(j+1, t) \leq 4J \int_0^t \textrm{d}t' f_x(j, t'),
\end{equation}
where $f_x(j, t) \equiv \lVert \mathcal{P}_{\geq j} A_x(t) \rVert$.

The proof in the following section will largely be devoted to deriving a substantially tighter result than Eq.~\eqref{eq:conventional_bound_local_inequality_v3} when random fields of the form in Eq.~\eqref{eq:local_field_distribution} are present.
Yet keep in mind that this does not invalidate Eq.~\eqref{eq:conventional_bound_local_inequality_v3} --- the conventional proof reproduced here continues to hold (it is simply not as tight a bound), and so we can (and will) use it in our subsequent analysis.

While we will not need it in what follows, let us show for completeness how the LR bound in Eq.~\eqref{eq:original_Lieb_Robinson_bound} follows straightforwardly from Eq.~\eqref{eq:conventional_bound_local_inequality_v3}.
First relate $f_x(y, t)$ to $f_x(y-1, t')$:
\begin{equation} \label{eq:conventional_bound_starting_iteration}
f_x(y, t) \leq 4J \int_0^t \textrm{d}t' f_x(y-1, t').
\end{equation}
Then relate $f_x(y-1, t')$ to $f_x(y-2, t'')$ analogously, then to $f_x(y-3, t''')$, and so on until reaching site $x$.
This gives a series of nested integrals:
\begin{equation} \label{eq:conventional_bound_repeated_iteration}
\begin{aligned}
f_x(y, t) &\leq (4J)^{|y-x|} \int_0^t \textrm{d}t_{y-1} \int_0^{t_{y-1}} \textrm{d}t_{y-2} \\
&\qquad \qquad \qquad \qquad \cdots \int_0^{t_{x+1}} \textrm{d}t_x f_x(x, t_x).
\end{aligned}
\end{equation}
Use that $f_x(x, t_x) \leq 2 \lVert A_x \rVert$ (Eq.~\eqref{eq:conventional_bound_technical_lemma}), and then the nested integrals can be evaluated:
\begin{equation} \label{eq:conventional_bound_final_result_v1}
f_x(y, t) \leq 2 \lVert A_x \rVert \frac{(4Jt)^{|y-x|}}{|y-x|!}.
\end{equation}
While this is already a reasonably simple expression, some further inequalities ($a! \geq (a/e)^a$ and $\log{a} \leq a-1$ for all $a$) reduce Eq.~\eqref{eq:conventional_bound_final_result_v1} to
\begin{equation} \label{eq:conventional_bound_final_result_v2}
f_x(y, t) \leq 2 \lVert A_x \rVert e^{4eJt - |y-x|},
\end{equation}
which is the claimed result.

As a final comment, note that when we insert $\mathcal{P}_{\geq j}$ between $\overline{\mathcal{L}}_{j,j+1} \overline{A}_x$ (to obtain Eq.~\eqref{eq:conventional_bound_interaction_equation_projected}), we could equally well insert $\mathcal{P}_{\geq k}$ for any $k \leq j$ --- any basis string that has the identity beyond $k \leq j$ is clearly annihilated by $\overline{\mathcal{L}}_{j,j+1}$ as well.
When the interactions are uniform, this would only give a weaker LR bound.
Yet when the interactions are \textit{disordered}, this can in fact yield significantly tighter bounds with potentially larger dynamical exponents~\cite{Baldwin2023Disordered}.
We will use the same approach in what follows.

\section{Proof} \label{sec:proof}

We are deriving an improved LR bound for Hamiltonians (acting on $N$ spins with $d$ states each) of the form
\begin{equation} \label{eq:general_Hamiltonian_form_repeat}
H(t) = \sum_{j=1}^{N-1} H_{j,j+1}(t) - \sum_{j=1}^N h_j \hat{s}_j,
\end{equation}
such that
\begin{equation} \label{eq:norm_bounds_repeat}
\big\lVert H_{j,j+1}(t) \big\rVert \leq J, \quad \; \; \big\lVert \partial_t H_{j,j+1}(t) \big\rVert \leq \zeta J, \quad \; \; \big\lVert \hat{s}_j \big\rVert \leq b,
\end{equation}
and the random fields $h_j$ are drawn independently from a distribution with the large-$h$ behavior
\begin{equation} \label{eq:local_field_distribution_repeat}
\textrm{Pr} \big[ |h_j| > h \big] \sim Ch^{-\alpha}.
\end{equation}
Note that the fields can have random signs (as they often do in empirical studies), and Eq.~\eqref{eq:local_field_distribution_repeat} refers only to how the magnitudes are distributed.
Label the eigenvalues of $\hat{s}_j$ by $e_j^a$ and the projectors onto the associated eigenspaces by $P_j^a$, where $a \in \{1, \cdots, d\}$.
We assume that each $\hat{s}_j$ is non-degenerate, specifically such that
\begin{equation} \label{eq:local_term_gap_bound}
\min_j \min_{a \neq c} \big| e_j^a - e_j^c \big| \geq \Delta > 0,
\end{equation}
for some constant $\Delta$ which is independent of $N$.

\begin{figure}[t]
\centering
\includegraphics[width=1.0\columnwidth]{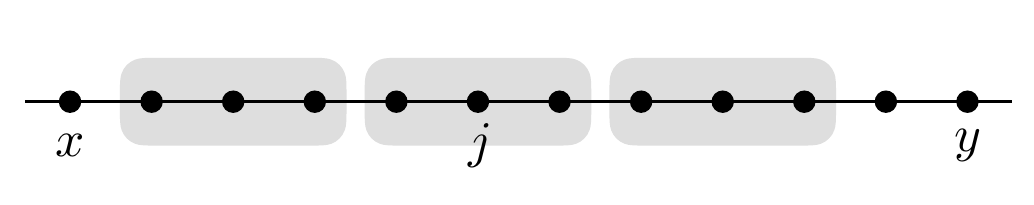}
\caption{Grouping of sites between $x$ and $y$ into blocks of three (gray rectangles). The first block begins at site $x+1$, and the final one or two sites are not included in a block if $|y-x|$ is not a multiple of three. In Secs.~\ref{subsec:tighter_local_relationship} and~\ref{subsec:lack_spreading_conserved_spins}, $j$ denotes the central site of a block.}
\label{fig:block_grouping}
\end{figure}

Define $r \equiv |y-x|$ for convenience.
Also define $h_0$ to be some fixed field strength such that each $|h_j|$ has probability at least 1/2 to be less than $h_0$, i.e.,
\begin{equation} \label{eq:local_field_typical_scale}
\textrm{Pr} \big[ |h_j| < h_0 \big] \equiv p \geq \frac{1}{2}.
\end{equation}
Group the sites between $x$ and $y$ into blocks of three sites each (ignoring the final one or two if $r$ is not a multiple of three) --- see Fig.~\ref{fig:block_grouping}.
Choose $\beta \in (0, 1/\alpha)$, and specifically consider blocks (called ``obstacles'') such that:
\begin{itemize}
\item The central site in the block --- call it site $j$ --- has a field larger than $r^{\beta}$ in magnitude.
\item The two peripheral sites in the block, sites $j-1$ and $j+1$, have fields less than $h_0$ in magnitude.
\end{itemize}
Denote the number of such obstacles by $n(r)$ --- one expects $n(r)$ to scale as $r/3$ times the probability of a block satisfying these two conditions (which according to Eqs.~\eqref{eq:local_field_distribution_repeat} and~\eqref{eq:local_field_typical_scale} is $Cp^2 r^{-\alpha \beta}$), and while we will confirm that this is indeed true, $n(r)$ is technically a random variable and so care is needed in order to take the limit properly.
For now, simply assume that $n(r) \geq 1$, i.e., that the realization of random fields is such that at least one obstacle exists between $x$ and $y$.

\subsection{Tighter local relationship} \label{subsec:tighter_local_relationship}

Here we derive an analogue to Eq.~\eqref{eq:conventional_bound_local_inequality_v3} from the conventional LR bound that has a much smaller prefactor due to the strong local field.
Choose an obstacle between $x$ and $y$, meaning that the central site $j$ of the block has $|h_j| > r^{\beta}$ while $|h_{j-1}| < h_0$ and $|h_{j+1}| < h_0$.
We begin as in Sec.~\ref{subsec:conventional_LR_bound}, by again making the interaction-picture transformation
\begin{equation} \label{eq:field_removal_transformation}
\overline{A}_x(t) \equiv e^{i \sum_{j=1}^N h_j \hat{s}_j t} A_x(t) e^{-i \sum_{j=1}^N h_j \hat{s}_j t}.
\end{equation}
Thus we again have
\begin{equation} \label{eq:field_removal_equation_of_motion}
\partial_t \overline{A}_x(t) = i \big[ \overline{H}(t), \overline{A}_x(t) \big],
\end{equation}
with the same Hamiltonian $\overline{H}$ as in Eq.~\eqref{eq:conventional_bound_field_removal_interactions}, but now write it slightly differently as
\begin{widetext}
\begin{equation} \label{eq:field_removal_full_Hamiltonian_v1}
\overline{H}(t) = \sum_{k \neq j-1, j} \overline{H}_{k,k+1}(t) + e^{i h_j \hat{s}_j t} \big( \overline{H}_{j-1,j}(t) + \overline{H}_{j,j+1}(t) \big) e^{-i h_j \hat{s}_j t},
\end{equation}
where $\overline{H}_{k,k+1}(t)$ is the same as in Sec.~\ref{subsec:conventional_LR_bound} but
\begin{equation} \label{eq:field_removal_individual_terms}
\overline{H}_{j-1,j}(t) \equiv e^{i h_{j-1} \hat{s}_{j-1} t} H_{j-1,j}(t) e^{-i h_{j-1} \hat{s}_{j-1} t}, \qquad \overline{H}_{j,j+1}(t) \equiv e^{i h_{j+1} \hat{s}_{j+1} t} H_{j,j+1}(t) e^{-i h_{j+1} \hat{s}_{j+1} t}.
\end{equation}
In other words, we write the dependence on $h_j$ explicitly rather than lumping it into $\overline{H}_{j-1,j}$ and $\overline{H}_{j,j+1}$.
Further insert resolutions of the identity via $\sum_a P_j^a = I$ (recall that $P_j^a$ projects onto spin $j$ being in state $a$):
\begin{equation} \label{eq:field_removal_full_Hamiltonian_v2}
\overline{H}(t) = \sum_{k \neq j-1, j} \overline{H}_{k,k+1}(t) + \sum_{a,c = 1}^d e^{i h_j (e_j^a - e_j^c) t} P_j^a \big( \overline{H}_{j-1,j}(t) + \overline{H}_{j,j+1}(t) \big) P_j^c.
\end{equation}

Write $\overline{H}(t) = \overline{H}_{\textrm{fix}}(t) + \overline{H}_{\textrm{osc}}(t)$, where $\overline{H}_{\textrm{fix}}(t)$ contains the terms that preserve the state of spin $j$ and $\overline{H}_{\textrm{osc}}(t)$ contains the terms that do not:
\begin{equation} \label{eq:Hamiltonian_fixed_terms}
\overline{H}_{\textrm{fix}}(t) \equiv \sum_{k \neq j-1, j} \overline{H}_{k,k+1}(t) + \sum_{a=1}^d P_j^a \big( \overline{H}_{j-1,j}(t) + \overline{H}_{j,j+1}(t) \big) P_j^a,
\end{equation}
\begin{equation} \label{eq:Hamiltonian_oscillating_terms}
\overline{H}_{\textrm{osc}}(t) \equiv \sum_{a \neq c} e^{i h_j (e_j^a - e_j^c) t} P_j^a \big( \overline{H}_{j-1,j}(t) + \overline{H}_{j,j+1}(t) \big) P_j^c.
\end{equation}
Note that we have the bound
\begin{equation} \label{eq:oscillating_term_magnitude_bound}
\big\lVert \overline{H}_{\textrm{osc}}(t) \big\rVert \leq \sum_{a \neq c} \big\lVert \overline{H}_{j-1,j}(t) + \overline{H}_{j,j+1}(t) \big\rVert \leq 2d(d-1)J.
\end{equation}
Passing to the interaction picture with respect to $\overline{\mathcal{U}}_{\textrm{fix}}$ gives, analogous to Eq.~\eqref{eq:conventional_bound_interaction_equation_differential},
\begin{equation} \label{eq:interaction_equation_differential}
\partial_t \Big( \overline{\mathcal{U}}_{\textrm{fix}}(t, 0)^{\dag} \overline{A}_x(t) \Big) = \overline{\mathcal{U}}_{\textrm{fix}}(t, 0)^{\dag} \overline{\mathcal{L}}_{\textrm{osc}}(t) \overline{A}_x(t),
\end{equation}
and thus, analogous to Eq.~\eqref{eq:conventional_bound_interaction_equation_integral},
\begin{equation} \label{eq:interaction_equation_integral}
\overline{A}_x(t) = \overline{\mathcal{U}}_{\textrm{fix}}(t, 0) A_x + \int_0^t \textrm{d}t' \overline{\mathcal{U}}_{\textrm{fix}}(t, t') \overline{\mathcal{L}}_{\textrm{osc}}(t') \overline{A}_x(t').
\end{equation}

Define for convenience $\overline{H}_B^{ac}(t) \equiv P_j^a (\overline{H}_{j-1,j}(t) + \overline{H}_{j,j+1}(t)) P_j^c$.
We have the bounds $\lVert \overline{H}_B^{ac}(t) \rVert \leq 2J$ and
\begin{equation} \label{eq:block_term_derivative_bound}
\begin{aligned}
\big\lVert \partial_t \overline{H}_B^{ac}(t) \big\rVert &= \Big\lVert P_j^a \Big( e^{i h_{j-1} \hat{s}_{j-1} t} \partial_t H_{j-1,j}(t) e^{-i h_{j-1} \hat{s}_{j-1} t} + i h_{j-1} \big[ \hat{s}_{j-1}, \overline{H}_{j-1,j}(t) \big] \Big) P_j^c \\
&\qquad + P_j^a \Big( e^{i h_{j+1} \hat{s}_{j+1} t} \partial_t H_{j,j+1}(t) e^{-i h_{j+1} \hat{s}_{j+1} t} + i h_{j+1} \big[ \hat{s}_{j+1}, \overline{H}_{j,j+1}(t) \big] \Big) P_j^c \Big\rVert \\
&\leq 2 \zeta J + 4h_0 bJ,
\end{aligned}
\end{equation}
where the final expression follows from Eq.~\eqref{eq:norm_bounds_repeat} and the fact that, by assumption, $|h_{j-1}| < h_0$ and $|h_{j+1}| < h_0$.
Note that we can write
\begin{equation} \label{eq:oscillating_Liouvillian_expression}
\overline{\mathcal{L}}_{\textrm{osc}}(t) = \sum_{a \neq c} e^{i h_j (e_j^a - e_j^c) t} \overline{\mathcal{L}}_B^{ac}(t).
\end{equation}

As the subscript suggests, $\overline{\mathcal{L}}_{\textrm{osc}}$ is rapidly oscillating in time --- recall that $|h_j| > r^{\beta} \gg 1$ at large $r$.
Thus integrate by parts to bring down a factor of $h_j$:
\begin{equation} \label{eq:interaction_equation_integration_by_parts}
\begin{aligned}
\overline{A}_x(t) - \overline{\mathcal{U}}_{\textrm{fix}}(t, 0) A_x &= \sum_{a \neq c} \int_0^t \textrm{d}t' e^{i h_j (e_j^a - e_j^c) t'} \overline{\mathcal{U}}_{\textrm{fix}}(t, t') \overline{\mathcal{L}}_B^{ac}(t') \overline{A}_x(t') \\
&= \sum_{a \neq c} \frac{1}{i h_j (e_j^a - e_j^c)} \bigg( e^{i h_j (e_j^a - e_j^c) t} \overline{\mathcal{L}}_B^{ac}(t) \overline{A}_x(t) - \overline{\mathcal{U}}_{\textrm{fix}}(t, 0) \overline{\mathcal{L}}_B^{ac}(0) A_x \bigg) \\
&\qquad - \sum_{a \neq c} \frac{1}{i h_j (e_j^a - e_j^c)} \int_0^t \textrm{d}t' e^{i h_j (e_j^a - e_j^c) t'} \bigg( \Big( \partial_{t'} \overline{\mathcal{U}}_{\textrm{fix}}(t, t') \Big) \overline{\mathcal{L}}_B^{ac}(t') \overline{A}_x(t') \\
&\qquad \qquad \qquad \qquad \qquad \qquad + \overline{\mathcal{U}}_{\textrm{fix}}(t, t') \Big( \partial_{t'} \overline{\mathcal{L}}_B^{ac}(t') \Big) \overline{A}_x(t') + \overline{\mathcal{U}}_{\textrm{fix}}(t, t') \overline{\mathcal{L}}_B^{ac}(t') \partial_{t'} \overline{A}_x(t') \bigg) \\
&= \sum_{a \neq c} \frac{1}{i h_j (e_j^a - e_j^c)} e^{i h_j (e_j^a - e_j^c) t} \overline{\mathcal{L}}_B^{ac}(t) \overline{A}_x(t) \\
&\qquad - \sum_{a \neq c} \frac{1}{i h_j (e_j^a - e_j^c)} \int_0^t \textrm{d}t' e^{i h_j (e_j^a - e_j^c) t'} \overline{\mathcal{U}}_{\textrm{fix}}(t, t') \bigg( -\overline{\mathcal{L}}_{\textrm{fix}}(t') \overline{\mathcal{L}}_B^{ac}(t') \overline{A}_x(t') \\
&\qquad \qquad \qquad \qquad \qquad \qquad \qquad \qquad \qquad + \Big( \partial_{t'} \overline{\mathcal{L}}_B^{ac}(t') \Big) \overline{A}_x(t') + \overline{\mathcal{L}}_B^{ac}(t') \overline{\mathcal{L}}(t') \overline{A}_x(t') \bigg).
\end{aligned}
\end{equation}
Note that the term $\overline{\mathcal{L}}_B^{ac}(0) A_x = i[\overline{H}_B^{ac}(0), A_x]$ vanishes since by assumption $x < j-1$ (we chose $j$ to be the central site in a block between $x$ and $y$ --- see Fig.~\ref{fig:block_grouping}).
Two of the remaining terms can be simplified via Jacobi's identity:
\begin{equation} \label{eq:interaction_equation_Jacobi_identity}
\begin{aligned}
&\overline{\mathcal{L}}_B^{ac}(t') \overline{\mathcal{L}}(t') \overline{A}_x(t') - \overline{\mathcal{L}}_{\textrm{fix}}(t') \overline{\mathcal{L}}_B^{ac}(t') \overline{A}_x(t') \\
&\qquad \qquad = \overline{\mathcal{L}}(t') \overline{\mathcal{L}}_B^{ac}(t') \overline{A}_x(t') - \Big[ \big[ \overline{H}_B^{ac}(t'), \overline{H}(t') \big], \overline{A}_x(t') \Big] - \overline{\mathcal{L}}_{\textrm{fix}}(t') \overline{\mathcal{L}}_B^{ac}(t') \overline{A}_x(t') \\
&\qquad \qquad = \overline{\mathcal{L}}_{\textrm{osc}}(t') \overline{\mathcal{L}}_B^{ac}(t') \overline{A}_x(t') - \Big[ \big[ \overline{H}_B^{ac}(t'), \overline{H}(t') \big], \overline{A}_x(t') \Big].
\end{aligned}
\end{equation}
Denote the superoperator that takes the commutator with $[\overline{H}_B^{ac}(t'), \overline{H}(t')]$ by $\mathcal{C}^{ac}(t')$, so that the final term of Eq.~\eqref{eq:interaction_equation_Jacobi_identity} can be written $-\mathcal{C}^{ac}(t') \overline{A}_x(t')$.
Note that since $\overline{H}_B^{ac}$ acts only within $[j-1, j+1]$, the only terms of $\overline{H}$ that contribute to the commutator are those within $[j-2, j+2]$ (four terms in total).
Thus $[\overline{H}_B^{ac}(t'), \overline{H}(t')]$ itself acts only within $[j-2, j+2]$, and we have the simple bound
\begin{equation} \label{eq:interaction_commutator_bound}
\Big\lVert \big[ \overline{H}_B^{ac}(t'), \overline{H}(t') \big] \Big\rVert \leq 2 (2J) (4J) = 16J^2.
\end{equation}
Returning to Eq.~\eqref{eq:interaction_equation_integration_by_parts}, we have the expression (which note is still an exact equality)
\begin{equation} \label{eq:interaction_equation_manipulated}
\begin{aligned}
&\overline{A}_x(t) - \overline{\mathcal{U}}_{\textrm{fix}}(t, 0) A_x = \sum_{a \neq c} \frac{1}{i h_j (e_j^a - e_j^c)} e^{i h_j (e_j^a - e_j^c) t} \overline{\mathcal{L}}_B^{ac}(t) \overline{A}_x(t) \\
&\qquad \qquad \qquad \qquad - \sum_{a \neq c} \frac{1}{i h_j (e_j^a - e_j^c)} \int_0^t \textrm{d}t' e^{i h_j (e_j^a - e_j^c) t'} \overline{\mathcal{U}}_{\textrm{fix}}(t, t') \Big( \overline{\mathcal{L}}_{\textrm{osc}}(t') \overline{\mathcal{L}}_B^{ac}(t') + \partial_{t'} \overline{\mathcal{L}}_B^{ac}(t') - \mathcal{C}^{ac}(t') \Big) \overline{A}_x(t').
\end{aligned}
\end{equation}

Now act on both sides with the projector $\mathcal{P}_{\geq j+1}$.
On the left-hand side, though it may not be immediately obvious, we show below that $\mathcal{P}_{\geq j+1} \overline{\mathcal{U}}_{\textrm{fix}}(t, 0) A_x = 0$.
On the right-hand side, every term involves acting on $\overline{A}_x(t')$ with a commutator, and every interaction term in such a commutator acts only near site $j$.
By the same reasoning as in Sec.~\ref{subsec:conventional_LR_bound}, we can replace $\overline{A}_x(t')$ by $\mathcal{P}_{\geq k'} \overline{A}_x(t')$ as long as site $k'$ lies to the left of the region where the interaction terms act.
With a view to later steps, choose $k$ to be the central site in the preceding obstacle (or, if none exists, set $k = x-1$) --- see Fig.~\ref{fig:projector_insertions}.
Note that $k \leq j-3$.
In the first term on the right-hand side, substitute $\mathcal{P}_{\geq k+2} \overline{A}_x(t)$ (allowed since $\overline{H}_B^{ac}$ acts no further to the left than site $j-1$).
In the second term, substitute $\mathcal{P}_{\geq k+1} \overline{A}_x(t')$ (allowed since the furthest left that any interaction acts is site $j-2$, coming from $\mathcal{C}^{ac}$).
Putting everything together, we have
\begin{equation} \label{eq:interaction_equation_projected}
\begin{aligned}
&\mathcal{P}_{\geq j+1} \overline{A}_x(t) = \sum_{a \neq c} \frac{1}{i h_j (e_j^a - e_j^c)} e^{i h_j (e_j^a - e_j^c) t} \mathcal{P}_{\geq j+1} \overline{\mathcal{L}}_B^{ac}(t) \mathcal{P}_{\geq k+2} \overline{A}_x(t) \\
&\qquad \qquad - \sum_{a \neq c} \frac{1}{i h_j (e_j^a - e_j^c)} \int_0^t \textrm{d}t' e^{i h_j (e_j^a - e_j^c) t'} \mathcal{P}_{\geq j+1} \overline{\mathcal{U}}_{\textrm{fix}}(t, t') \Big( \overline{\mathcal{L}}_{\textrm{osc}}(t') \overline{\mathcal{L}}_B^{ac}(t') + \partial_{t'} \overline{\mathcal{L}}_B^{ac}(t') - \mathcal{C}^{ac}(t') \Big) \mathcal{P}_{\geq k+1} \overline{A}_x(t').
\end{aligned}
\end{equation}

Now take norms.
By definition, we have $|h_j| > r^{\beta}$ and $|e_j^a - e_j^c| \geq \Delta$.
We have established bounds on all interactions entering the right-hand side: Eqs.~\eqref{eq:oscillating_term_magnitude_bound},~\eqref{eq:block_term_derivative_bound}, and~\eqref{eq:interaction_commutator_bound}.
We can use Eq.~\eqref{eq:conventional_bound_technical_lemma} to bound $\lVert \mathcal{P}_{\geq j+1} \cdots \rVert$ by $2 \lVert \cdots \rVert$.
Thus we arrive at the result
\begin{equation} \label{eq:strong_field_local_relationship_v1}
f_x(j+1, t) \leq \frac{8d(d-1)J}{r^{\beta} \Delta} \bigg[ f_x(k+2, t) + \big( \zeta + 2h_0 b + 4(d^2 - d + 2)J \big) \int_0^t \textrm{d}t' f_x(k+1, t') \bigg],
\end{equation}
\end{widetext}
where $f_x(j, t) \equiv \lVert \mathcal{P}_{\geq j} \overline{A}_x(t) \rVert = \lVert \mathcal{P}_{\geq j} A_x(t) \rVert$.
Lastly use Eq.~\eqref{eq:conventional_bound_local_inequality_v3} from the conventional bound (which remains a valid result) to relate $f_x(k+2, t)$ to $f_x(k+1, t')$.
We have
\begin{equation} \label{eq:strong_field_local_relationship_v2}
f_x(j+1, t) \leq \frac{K}{r^{\beta}} \int_0^t \textrm{d}t' f_x(k+1, t'),
\end{equation}
where $K \equiv 8d(d-1)J [\zeta + 2h_0 b + 4(d^2 - d + 3)J]/\Delta$.

\begin{figure}
\centering
\includegraphics[width=0.95\columnwidth]{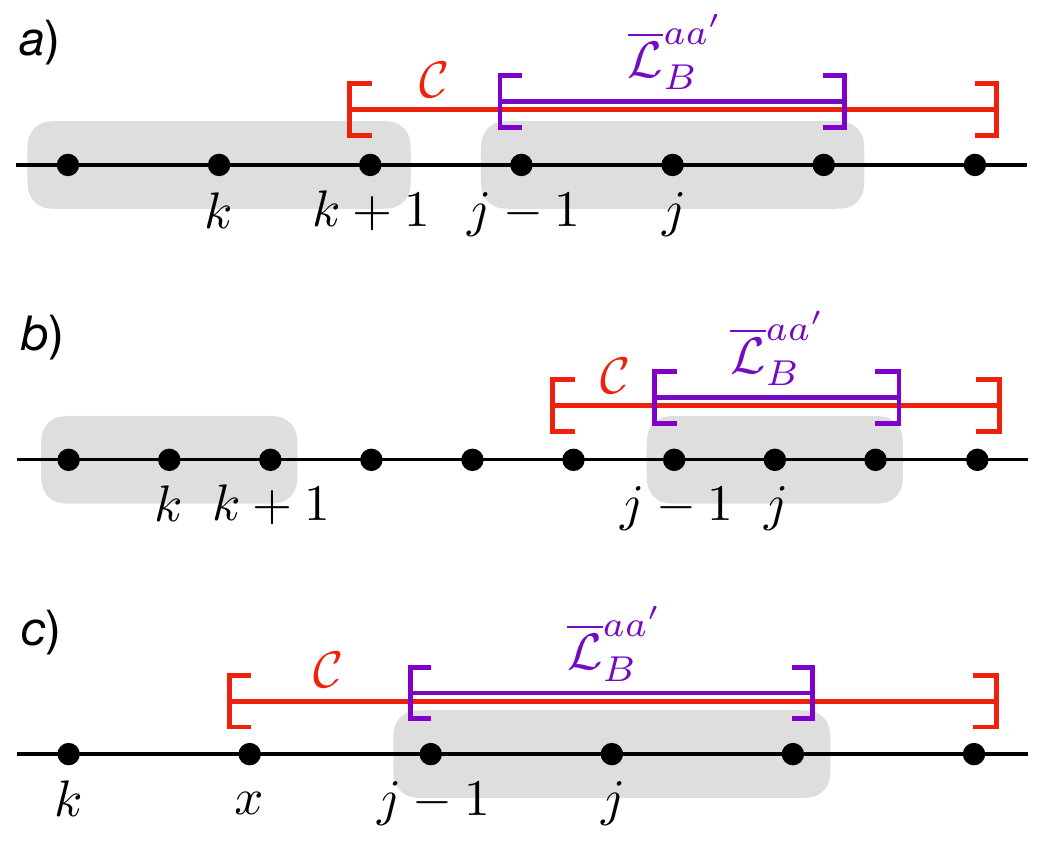}
\caption{Illustration of how to choose site $k$ when inserting projectors into Eq.~\eqref{eq:interaction_equation_manipulated}. Gray rectangles indicate ``obstacles'' as defined in the text. The purple interval indicates the range of sites over which $\overline{\mathcal{L}}_B^{ac}$ acts, and the red interval indicates the range over which $\mathcal{C}$ acts. a) When the block neighboring that of $j$ is itself an obstacle, choose $k$ to be the central site as shown. Then $\overline{\mathcal{L}}_B^{ac}$ acts only in the region $\geq k+2$, and $\mathcal{C}$ acts only in the region $\geq k+1$. b) If the preceding obstacle is further to the left, still take $k$ to be its central site. It is still true that $\overline{\mathcal{L}}_B^{ac}$ acts within $\geq k+2$ and $\mathcal{C}$ acts within $\geq k+1$. c) If there is no preceding obstacle, take $k = x-1$, so that $\overline{\mathcal{L}}_B^{ac}$ still acts within $\geq k+2$ and $\mathcal{C}$ within $\geq k+1$.}
\label{fig:projector_insertions}
\end{figure}

Eq.~\eqref{eq:strong_field_local_relationship_v2} is the desired analogue to Eq.~\eqref{eq:conventional_bound_local_inequality_v3}.
It relates the weight of $A_x(t)$ past an obstacle to the weight past the preceding obstacle, and thus we can iterate~\cite{iteration_note} to obtain a series of nested integrals as in Eq.~\eqref{eq:conventional_bound_repeated_iteration}:
\begin{equation} \label{eq:strong_field_repeated_iteration}
\begin{aligned}
f_x(y, t) &\leq 2 \cdot \frac{K}{r^{\beta}} \int_0^t \textrm{d}t_{j_2} \frac{K}{r^{\beta}} \int_0^{t_{j_2}} \textrm{d}t_{j_3} \\
&\qquad \qquad \qquad \cdots \frac{K}{r^{\beta}} \int_0^{t_{j_{n-1}}} \textrm{d}t_x f_x(x, t_x).
\end{aligned}
\end{equation}
The number of nested integrals is the number of obstacles between $x$ and $y$, which we are denoting by $n(r)$, and thus this evaluates to (using $f_x(x, t_x) \leq 2 \lVert A_x \rVert$)
\begin{equation} \label{eq:strong_field_LR_bound_implicit_v1}
f_x(y, t) \leq 4 \lVert A_x \rVert \frac{1}{n(r)!} \left( \frac{Kt}{r^{\beta}} \right)^{n(r)}.
\end{equation}
The same inequalities used to reduce Eq.~\eqref{eq:conventional_bound_final_result_v1} to Eq.~\eqref{eq:conventional_bound_final_result_v2} then give
\begin{equation} \label{eq:strong_field_LR_bound_implicit_v2}
f_x(y, t) \leq 4 \lVert A_x \rVert \exp{\left[ \frac{eKt}{r^{\beta}} - n(r) \right]}.
\end{equation}

The final step is to determine the scaling of $n(r)$ to sufficient accuracy and with sufficient probability (note that $n(r)$ is technically a random variable that depends on the specific realization of random fields).
First, however, we establish that $\mathcal{P}_{\geq j+1} \overline{\mathcal{U}}_{\textrm{fix}}(t, 0) A_x$ is indeed 0, as claimed below Eq.~\eqref{eq:interaction_equation_manipulated}.

\subsection{Lack of spreading past conserved spins} \label{subsec:lack_spreading_conserved_spins}

In words, we are proving that evolution under $\overline{H}_{\textrm{fix}}$ alone (Eq.~\eqref{eq:Hamiltonian_fixed_terms}) cannot cause an operator initially acting to the left of site $j$ to spread to the right of site $j$.
This is due to the fact that $\overline{H}_{\textrm{fix}}$ preserves the state of spin $j$, i.e., is diagonal in the basis of $\hat{s}_j$ eigenstates.

Write $\overline{\mathcal{U}}_{\textrm{fix}}(t, 0) A_x \equiv A_{\textrm{fix}}(t)$ for convenience.
We show that $\mathcal{P}_{\geq j+1} A_{\textrm{fix}}(t) = 0$ by choosing a sufficiently small $\Delta t$ and considering the first-order approximation to $A_{\textrm{fix}}(t)$, i.e., the operator $A_{\textrm{fix}}^m$ (meant to approximate $A_{\textrm{fix}}(m \Delta t)$) whose ``dynamics'' is given by
\begin{equation} \label{eq:first_order_Heisenberg_equation}
A_{\textrm{fix}}^{m+1} = A_{\textrm{fix}}^m + i \Delta t \big[ \overline{H}_{\textrm{fix}}(m \Delta t), A_{\textrm{fix}}^m \big], \qquad A_{\textrm{fix}}^0 = A_x.
\end{equation}
We prove by induction that $\mathcal{P}_{\geq j+1} A_{\textrm{fix}}^m = 0$ for all $m$.
Denoting $m_t \equiv t/\Delta t$, we then have
\begin{equation} \label{eq:first_order_exact_projection_relationship}
\begin{aligned}
&\big\lVert \mathcal{P}_{\geq j+1} A_{\textrm{fix}}(t) \big\rVert \\
&\qquad \leq \big\lVert \mathcal{P}_{\geq j+1} \big( A_{\textrm{fix}}(t) - A_{\textrm{fix}}^{m_t} \big) \big\rVert + \big\lVert \mathcal{P}_{\geq j+1} A_{\textrm{fix}}^{m_t} \big\rVert \\
&\qquad \leq 2 \big\lVert A_{\textrm{fix}}(t) - A_{\textrm{fix}}^{m_t} \big\rVert.
\end{aligned}
\end{equation}
and since $\lVert A_{\textrm{fix}}(t) - A_{\textrm{fix}}^{m_t} \rVert$ can be made arbitrarily small by taking $\Delta t \rightarrow 0$~\cite{Iserles2008}, we have that $\lVert \mathcal{P}_{\geq j+1} A_{\textrm{fix}}(t) \rVert = 0$ as well.

Clearly $\mathcal{P}_{\geq j+1} A_{\textrm{fix}}^0 = \mathcal{P}_{\geq j+1} A_x = 0$.
For the inductive step, assume that $A_{\textrm{fix}}^m$ does not act past site $j$ and that it is furthermore of the form $\sum_a P_j^a A_{\textrm{fix}}^m P_j^a$, i.e., diagonal on site $j$.
We prove that the same is true of $[\overline{H}_{\textrm{fix}}(m \Delta t), A_{\textrm{fix}}^m]$.
Write $\overline{H}_{\textrm{fix}}$ as (suppressing time arguments for brevity)
\begin{equation} \label{eq:fixed_Hamiltonian_compact_form}
\overline{H}_{\textrm{fix}} = \overline{H}_{\leq j-1} + \sum_a P_j^a \big( \overline{H}_{j-1,j} + \overline{H}_{j,j+1} \big) P_j^a + \overline{H}_{\geq j+1},
\end{equation}
where $\overline{H}_{\leq j-1}$ denotes all terms acting entirely to the left of site $j$, and analogously for $\overline{H}_{\geq j+1}$.
We have, using that $\overline{H}_{\geq j+1}$ commutes with $A_{\textrm{fix}}^m$ by assumption,
\begin{widetext}
\begin{equation} \label{eq:fixed_Hamiltonian_full_commutator}
\begin{aligned}
\big[ \overline{H}_{\textrm{fix}}, A_{\textrm{fix}}^m \big] &= \sum_a \big[ \overline{H}_{\textrm{fix}}, P_j^a A_{\textrm{fix}}^m P_j^a \big] \\
&= \sum_a \big[ \overline{H}_{\leq j-1}, P_j^a A_{\textrm{fix}}^m P_j^a \big] + \sum_{ac} \big[ P_j^c \big( \overline{H}_{j-1,j} + \overline{H}_{j,j+1} \big) P_j^c, P_j^a A_{\textrm{fix}}^m P_j^a \big] \\
&= \sum_a \big[ \overline{H}_{\leq j-1}, P_j^a A_{\textrm{fix}}^m P_j^a \big] + \sum_a \big[ P_j^a \overline{H}_{j-1,j} P_j^a, P_j^a A_{\textrm{fix}}^m P_j^a \big] + \sum_a \big[ P_j^a \overline{H}_{j,j+1} P_j^a, P_j^a A_{\textrm{fix}}^m P_j^a \big].
\end{aligned}
\end{equation}
\end{widetext}

For the first term on the bottom line, since $\overline{H}_{\leq j-1}$ commutes with $P_j^a$, we can write it as $\sum_a P_j^a [\overline{H}_{\leq j-1}, A_{\textrm{fix}}^m] P_j^a$, which is indeed an operator that acts no further than site $j$ and is diagonal on $j$.
The second term is also of the form $\sum_a P_j^a \cdots P_j^a$, acting no further than site $j$.
The third term in fact vanishes: $j$ is the only site in common on which $P_j^a \overline{H}_{j,j+1} P_j^a$ and $P_j^a A_{\textrm{fix}}^m P_j^a$ act, but they are both diagonal on that site and so trivially commute.
More specifically, expand $P_j^a A_{\textrm{fix}}^m P_j^a$ as a linear combination of basis strings (see Sec.~\ref{subsec:definitions}):
\begin{equation} \label{eq:fixed_operator_basis_expansion}
\begin{aligned}
&P_j^a A_{\textrm{fix}}^m P_j^a \\
&\qquad = {\sum_{\nu}}' c'_{\nu} \big( \cdots \otimes X_{j-1}^{\nu_{j-1}} \otimes P_j^a X_j^{\nu_j} P_j^a \otimes I_{j+1} \otimes \cdots \big),
\end{aligned}
\end{equation}
where the prime indicates that only strings having the identity on all sites past $j$ are to be included in the sum.
Similarly, $P_j^a \overline{H}_{j,j+1} P_j^a$ can be written
\begin{equation} \label{eq:fixed_Hamiltonian_term_basis_expansion}
\begin{aligned}
&P_j^a \overline{H}_{j,j+1} P_j^a \\
&\qquad = {\sum_{\upsilon}}'' c''_{\upsilon} \big( \cdots \otimes I_{j-1} \otimes P_j^a X_j^{\upsilon_j} P_j^a \otimes X_{j+1}^{\upsilon_{j+1}} \otimes \cdots \big),
\end{aligned}
\end{equation}
where the double-prime indicates summing only over strings acting solely on sites $j$ and $j+1$.
Each individual term in Eq.~\eqref{eq:fixed_operator_basis_expansion} commutes with each individual term in Eq.~\eqref{eq:fixed_Hamiltonian_term_basis_expansion}, thus $P_j^a A_{\textrm{fix}}^m P_j^a$ commutes with $P_j^a \overline{H}_{j,j+1} P_j^a$ as a whole.

Since the third term of Eq.~\eqref{eq:fixed_Hamiltonian_full_commutator} vanishes, $A_{\textrm{fix}}^m$ and $[\overline{H}_{\textrm{fix}}(m \Delta t), A_{\textrm{fix}}^m]$ thus both are operators that act no further than site $j$ and are diagonal on $j$, meaning $A_{\textrm{fix}}^{m+1}$ is as well (Eq.~\eqref{eq:first_order_Heisenberg_equation}).
By induction, the same is true for $A_{\textrm{fix}}^{m_t}$, i.e., $\mathcal{P}_{\geq j+1} A_{\textrm{fix}}^{m_t} = 0$ as claimed.

\subsection{Counting obstacles} \label{subsec:quantifying_targeted_blocks}

The remainder of the proof is entirely analogous to that in Ref.~\cite{Baldwin2023Disordered}, but we include the details for completeness.
In this subsection, we prove that with probability 1,
\begin{equation} \label{eq:targeted_block_number}
\lim_{r \rightarrow \infty} \frac{n(r)}{\mathbb{E} n(r)} = 1,
\end{equation}
where $\mathbb{E} n(r)$ is the average number of obstacles between $x$ and $y$.
The probability of any given block being an obstacle, denoted $\mu(r)$, is
\begin{equation} \label{eq:single_block_target_probability}
\mu(r) = \textrm{Pr} \big[ |h_{j-1}| < h_0 \big] \cdot \textrm{Pr} \big[ |h_j| > r^{\beta} \big] \cdot \textrm{Pr} \big[ |h_{j+1}| < h_0 \big],
\end{equation}
and from Eqs.~\eqref{eq:local_field_distribution_repeat} and~\eqref{eq:local_field_typical_scale}, $\mu(r) \sim Cp^2 r^{-\alpha \beta}$ at large $r$.
Since the total number of blocks between $x$ and $y$ goes as $r/3$, we have that
\begin{equation} \label{eq:targeted_block_average_number}
\mathbb{E} n(r) \sim \frac{C p^2 r^{1 - \alpha \beta}}{3}.
\end{equation}
In the following subsection, we use this limiting behavior of $n(r)$ (which again holds with probability 1) in Eq.~\eqref{eq:strong_field_LR_bound_implicit_v2} to obtain the final LR bound.

The proof of Eq.~\eqref{eq:targeted_block_number} is an application of the Borel-Cantelli lemma~\cite{Rosenthal2006}: for any infinite sequence of events $\{ E_r \}_{r=1}^{\infty}$, if the sum $\sum_r \textrm{Pr}[E_r]$ is finite, then with probability 1, only a finite number of the events will occur.
In our case, choose some $\epsilon > 0$ and take $E_r$ to be the event that $n(r) / \mathbb{E} n(r)$ lies outside the interval $(1-\epsilon, 1+\epsilon)$.
We will show that $\sum_r \textrm{Pr}[E_r]$ is finite, and thus with probability 1, $n(r) / \mathbb{E} n(r)$ deviates from $(1-\epsilon, 1+\epsilon)$ for only a finite number of $r$.
In other words, there is a distance $R(\epsilon)$ past which $n(r) / \mathbb{E} n(r) \in (1-\epsilon, 1+\epsilon)$.
Since the intersection of a countable number of probability-1 events continues to have probability 1~\cite{Rosenthal2006}, this means that the following statement holds with certainty: for any rational $\epsilon > 0$, there exists $R(\epsilon)$ such that $n(r) / \mathbb{E} n(r) \in (1-\epsilon, 1+\epsilon)$ for all $r > R(\epsilon)$.
This statement is precisely the definition of the limit in Eq.~\eqref{eq:targeted_block_number}.

We can calculate $\textrm{Pr}[E_r]$ directly.
Since there are $\lfloor r/3 \rfloor$ blocks between $x$ and $y$, where $\lfloor \cdots \rfloor$ denotes the integer part, and each has probability $\mu(r)$ (Eq.~\eqref{eq:single_block_target_probability}) of being an obstacle, the probability of there being $m$ obstacles is
\begin{equation} \label{eq:specific_target_number_probability}
\textrm{Pr} \big[ n(r) = m \big] = \binom{\lfloor r/3 \rfloor}{m} \mu(r)^m \big( 1 - \mu(r) \big)^{\lfloor r/3 \rfloor - m}.
\end{equation}
Then $\textrm{Pr}[E_r]$ is the sum over all $m$ outside the interval $(\mathbb{E} n(r) (1 - \epsilon), \mathbb{E} n(r) (1 + \epsilon))$:
\begin{widetext}
\begin{equation} \label{eq:target_number_deviation_probability}
\begin{aligned}
\textrm{Pr} \Big[ \frac{n(r)}{\mathbb{E} n(r)} \not \in (1 - \epsilon, 1 + \epsilon) \Big] &= \sum_{m \not \in (\mathbb{E} n(r) (1 - \epsilon), \mathbb{E} n(r) (1 + \epsilon))} \binom{\lfloor r/3 \rfloor}{m} \mu(r)^m \big( 1 - \mu(r) \big)^{\lfloor r/3 \rfloor - m} \\
&\leq \sum_{m \not \in (\mathbb{E} n(r) (1 - \epsilon), \mathbb{E} n(r) (1 + \epsilon))} \exp{\left[ m \log{\frac{\lfloor r/3 \rfloor \mu(r)}{m}} + \big( \lfloor r/3 \rfloor - m \big) \log{\frac{\lfloor r/3 \rfloor (1 - \mu(r))}{\lfloor r/3 \rfloor - m}} \right]} \\
&\leq \lfloor r/3 \rfloor \exp{\left[ -(1 + \epsilon) \lfloor r/3 \rfloor \mu(r) \log{(1 + \epsilon)} - \big( 1 - (1 + \epsilon) \mu(r) \big) \lfloor r/3 \rfloor \log{\frac{1 - (1 + \epsilon) \mu(r)}{1 - \mu(r)}} \right]},
\end{aligned}
\end{equation}
\end{widetext}
where the second line follows from Stirling’s formula (as an inequality) and the third line follows from replacing the summand by its maximum over allowed $m$ (noting that $\mathbb{E} n(r) = \lfloor r/3 \rfloor \mu(r)$).
Since $\mu(r) \sim Cp^2 r^{-\alpha \beta}$ at large $r$, one can confirm that the final line scales as $r \exp{[-Dr^{1 - \alpha \beta}]}$ with positive constant $D$.
This is a stretched exponential for any $\beta < 1/\alpha$, hence summable, and thus Eq.~\eqref{eq:targeted_block_number} is proven.

\subsection{Finishing the proof} \label{subsec:finishing_proof}

From Eqs.~\eqref{eq:targeted_block_number} and~\eqref{eq:targeted_block_average_number}, there is a $\beta$-dependent but finite distance $R(\beta)$ past which
\begin{equation} \label{eq:targeted_block_number_lower_bound}
n(r) \geq \frac{1}{2} \cdot \frac{Cp^2 r^{1 - \alpha \beta}}{3}.
\end{equation}
Returning to Eq.~\eqref{eq:strong_field_LR_bound_implicit_v2}, this gives
\begin{equation} \label{eq:strong_field_LR_bound_explicit}
\begin{aligned}
f_x(y, t) &\leq 4 \lVert A_x \rVert \exp{\left[ \frac{eKt}{r^{\beta}} - \frac{Cp^2 r^{1 - \alpha \beta}}{6} \right]} \\
&= 4 \lVert A_x \rVert \exp{\left[ -\frac{Cp^2 r^{1 - \alpha \beta}}{6} \left( 1 - \frac{vt}{r^{1 + (1 - \alpha) \beta}} \right) \right]},
\end{aligned}
\end{equation}
where $v \equiv 6eK/Cp^2$.
Eq.~\eqref{eq:strong_field_LR_bound_explicit} holds with probability 1 for any choice of $\beta \in (0, 1/\alpha)$ and at all distances greater than $R(\beta)$.
Note that $R(\beta)$ does depend on the specific realization of random fields, but it is at least guaranteed to be finite (and one could straightforwardly upper-bound the probability of $R(\beta)$ exceeding a given value).

The form in the bottom line shows that this LR bound has dynamical exponent $z = 1 + (1 - \alpha) \beta$: it defines a ``front'' given by the spacetime curve $vt = r^z$ such that (at large $r$) the LR bound is extremely small for $vt < r^z$ but extremely large (meaning a rather useless bound) for $vt > r^z$.

Suppose $\alpha > 1$.
Then $z = 1 + (1 - \alpha) \beta < 1$.
While such a bound is not incorrect, it is not useful since the conventional LR bound already has a dynamical exponent of 1.
Thus we will not consider $\alpha > 1$ further.

The case $\alpha < 1$ is more interesting.
The tightest LR bound has the largest dynamical exponent, and we can make $z$ arbitrarily close to $1/\alpha$ by taking $\beta \rightarrow 1/\alpha$ (although we cannot set $\beta = 1/\alpha$ exactly --- see Sec.~\ref{subsec:quantifying_targeted_blocks}).
This means that not only do we have an LR bound with dynamical exponent $z$ for any $z < 1/\alpha$, but the ``generalized velocity'' $v$ in such a bound can be made arbitrarily small: having set $z < 1/\alpha$, choose any $\beta$ such that $z < 1 + (1 - \alpha) \beta$, and then regardless of the value of $v$, the spacetime curve $vt = r^z$ will ultimately lie outside the front of the LR bound derived from $\beta$.

To be more quantitative, set $vt = r^z$ in Eq.~\eqref{eq:strong_field_LR_bound_explicit}:
\begin{equation} \label{eq:strong_field_LR_bound_minimized_v1}
f_x(y, t) \leq 4 \lVert A_x \rVert \exp{\left[ \frac{eK}{v} r^{z - \beta} - \frac{Cp^2 r^{1 - \alpha \beta}}{6} \right]}.
\end{equation}
Since $\beta$ is such that $z - \beta < 1 - \alpha \beta$, the second term in the exponent grows asymptotically faster than the first.
Thus for $r$ greater than a certain distance $R'$, we can bound
\begin{equation} \label{eq:strong_field_LR_bound_minimized_v2}
f_x(y, t) \leq 4 \lVert A_x \rVert \exp{\left[ -\frac{Cp^2 r^{1 - \alpha \beta}}{12} \right]}.
\end{equation}
The bound is thus stretched-exponential, with exponent $\gamma = 1 - \alpha \beta$.
The tightest bound (largest $\gamma$) comes from the smallest allowed $\beta$, i.e., taking $\beta \rightarrow (z - 1)/(1 - \alpha)$.
Then $\gamma$ becomes arbitrarily close to $(1 - \alpha z)/(1 - \alpha)$.
This result holds for any $z \geq 1$ (since we must have $\beta > 0$) --- for larger distances, we already know from the conventional LR bound that $f_x(y, t)$ is bounded by a regular exponential for all $r > 4eJt$.

To summarize (for $\alpha < 1$), there is a critical dynamical exponent given by $z_c = 1/\alpha$.
For $vt = |y-x|^z$ with any $z \in (1, z_c)$ and any $v > 0$, we have the LR bound
\begin{equation} \label{eq:strong_field_LR_bound_final}
f_x(y, t) \leq 4 \lVert A_x \rVert e^{-C' |y-x|^{\gamma}},
\end{equation}
for any $\gamma < (1 - \alpha z)/(1 - \alpha)$, obtained by setting $\beta = (1 - \gamma)/\alpha$ in our analysis.
The constant $C'$ depends only on the distribution of random fields and nothing else.
Eq.~\eqref{eq:strong_field_LR_bound_final} holds with probability 1 for all sites past a certain (random but finite) distance.

Before concluding, let us make two final comments.
First, one may wonder what can be said when $z = z_c$ exactly.
The conventional LR bound, after all, not only proves that super-ballistic operator growth is impossible but also identifies a finite maximum velocity $v_{\textrm{LR}}$.
In our case, writing $vt = r^{1/\alpha}$, we have been unable to establish an upper bound on the generalized velocity $v$ (and we suspect that it may be impossible to do so, given the results in Ref.~\cite{Baldwin2023Disordered}).
However, we can at least prove that there is a \textit{subsequence} of sites on which it is impossible for an operator to spread with \textit{any} non-zero $v$.
Fix $v > 0$ and $\epsilon > 0$, and consider every third site starting from the site at distance $R$, i.e., $y \in \{ x + R, x + R + 3, x + R + 6, \cdots \}$.
Exactly as in Ref.~\cite{Baldwin2023Disordered}, one can directly show that the probability is 0 of all such $y$ having either $|h_{y-1}| \geq h_0$, $|h_{y+1}| \geq h_0$, or $|h_y| \leq Kr^{1/\alpha}/v \epsilon$.
Thus the negation of the statement has probability 1, namely that for any distance $R$, there is a site $y$ beyond that distance with $|h_{y-1}| < h_0$, $|h_{y+1}| < h_0$, and $|h_y| > Kr^{1/\alpha}/v \epsilon$.
Eq.~\eqref{eq:strong_field_local_relationship_v2} thus applies to that site:
\begin{equation} \label{eq:very_strong_field_local_relationship}
f_x(y+1, t) \leq \frac{K}{|h_y|} \int_0^t \textrm{d}t' f_x(y-2, t) \leq 2 \lVert A_x \rVert \frac{Kt}{|h_y|}.
\end{equation}
Setting $vt = r^{1/\alpha}$ and with $|h_y| > Kr^{1/\alpha}/v \epsilon$, this gives $f_x(y+1, t) \leq 2 \lVert A_x \rVert \epsilon$.
This bound holds on an infinite number of sites (as there is always another beyond the current distance), for any $\epsilon$ and regardless of how small $v$ is, which proves the claim --- there is a subsequence $\{ y_i \}$ onto which no operator can spread with $O(1)$ amplitude in a time proportional to $|y_i - x|^{1/\alpha}$.

Second, one may wonder what can be said about the case $\alpha = 1$.
Our main results do not add anything to the conventional LR bound, since in this case they only establish that no operator can spread faster than ballistically (as is already known).
However, the observation of the preceding paragraph continues to hold: there is a subsequence of sites onto which no operator can spread with any finite velocity, however small.
Since the conventional bound only identifies a non-zero speed limit $v_{\textrm{LR}}$ for all sites, this statement is already a non-trivial improvement.
Furthermore, the larger question is whether our analysis can be expanded upon to derive even tighter dynamical exponents, perhaps ruling out ballistic motion for $\alpha \geq 1$ as well.
This remains to be seen, as discussed below.

\section{Conclusion} \label{sec:conclusion}

We have proven that in nearly any spin chain with power-law-distributed random fields, namely such that $\textrm{Pr}[|h_j| > h] \sim h^{-\alpha}$ at large $h$, there is an improved LR bound which guarantees that no operator can grow across a distance $r$ in time less than $O(r^{1/\alpha})$ when $\alpha < 1$.
See Eq.~\eqref{eq:strong_field_LR_bound_final} for the precise statement of the result, and see Eqs.~\eqref{eq:general_Hamiltonian_form} through~\eqref{eq:local_field_distribution} for the conditions under which it holds.
Note that this result proves the impossibility of any ballistic wave propagation for $\alpha < 1$ and the impossibility of any diffusive transport for $\alpha < 1/2$.
In fact, any finite dynamical exponent is ruled out at sufficiently small $\alpha$.

Note that the class of interactions to which our result applies is slightly more restrictive than that of the conventional LR bound: we require that the interactions be differentiable in time with bounded derivatives in addition to bounded norms, whereas the conventional case assumes only bounded norms.
We do not know whether this is the most general class of interactions for which an improved LR bound can be derived, but it is clear that some restriction is needed~\cite{Alexey_note}.
As explicitly used in our analysis, given a Hamiltonian with fields such as Eq.~\eqref{eq:general_Hamiltonian_form}, one can always make an interaction-picture transformation that removes the fields (see Eqs.~\eqref{eq:conventional_bound_field_removal_transformation} through~\eqref{eq:conventional_bound_field_removal_interactions}) --- the transformed interactions have different time-derivatives but the same norm since the transformation is unitary.
Conversely, given a Hamiltonian \textit{without} fields, one can always make the inverse transformation to introduce them.
Thus if it were possible to derive a sub-ballistic LR bound at strong fields which assumed nothing about the interactions beyond having bounded norm, that bound would in fact apply to all Hamiltonians without fields as well.
Since there certainly are field-less Hamiltonians that cause ballistic operator growth, such a general sub-ballistic bound must be impossible.

The most immediate open question is whether our result is tight, i.e., whether there exists a Hamiltonian consistent with Eqs.~\eqref{eq:general_Hamiltonian_form} through~\eqref{eq:local_field_distribution} for which some operator does grow with dynamical exponent $1/\alpha$.
Note that ``tightness'' here does not imply that every operator evolving under every such Hamiltonian must have this dynamical exponent, but rather merely a single example in which the bound is saturated --- even if our LR bound is found to be tight, there will certainly be numerous operators which grow much slower under specific Hamiltonians.

In the random-bond case, Ref.~\cite{Baldwin2023Disordered} shows that the LR bound derived there is indeed tight --- the Hamiltonian that saturates the bound amounts to simply applying a series of SWAP gates.
Such a straightforward construction does not apply to the random-field case for two reasons.
First, designing a high-fidelity SWAP gate in the presence of broadly distributed random fields is non-trivial and very well may be impossible.
Second, since we require the Hamiltonian to have bounded time-derivative, any such SWAP gate would have to be turned on smoothly (or at least differentiably), and the fields will induce additional complicated dynamics during that process (note that we are not allowed to turn the fields on and off).
Thus we leave the question of the tightness of our bound for future work.

As mentioned in the introduction, the intuition for our result derives from the strong-disorder renormalization group (RG) --- strong fields can amount to effective weak interactions under a suitable transformation.
It is interesting to note that this idea has already been explored in the context of MBL~\cite{Vosk2013ManyBody,Pekker2014HilbertGlass}.
Those works apply the RG transformation repeatedly (albeit with uncontrolled approximations) so as to study the fixed point for the effective fields and interactions.
Our technique, by contrast, is analogous to stopping after only a single transformation.
It would be quite interesting if this analysis could be incorporated into an iterative scheme to obtain even stronger results, particularly for random fields with milder distributions.

One final open question is whether and how our dynamical results connect to eigenstate properties, at least for time-independent Hamiltonians.
Many-body eigenstates have featured prominently in studies of MBL, in part because they are dramatic counter-examples to the ``eigenstate thermalization hypothesis'' (ETH) that is expected to hold for generic many-body systems~\cite{DAlessio2016From}.
It would be worth investigating the eigenstate properties in our regime $\alpha < 1$ for multiple reasons: to determine whether it is possible to obey ETH despite the sub-ballistic LR bounds, to see whether one generically finds signatures of MBL (recall that our result holds even when typical field values are much \textit{weaker} than the interaction strengths), and to look for new eigenstate features that might lie in between those of ETH and MBL.

\section{Acknowledgements} \label{sec:acknowledgements}

It is a pleasure to thank Anushya Chandran, Pieter W.\ Claeys, Alexey V.\ Gorshkov, Christopher R.\ Laumann, Antonello Scardicchio, and Daniele Toniolo for helpful comments, as well as Francois Huveneers for valuable discussion of their recent work~\cite{DeRoeck2024Absence}.
This work was performed with start-up funds from the physics department at Michigan State University.

\bibliography{biblio}

\begin{thebibliography}{94}
\bibitem [{tim()}]{time_derivative_note}%
  \BibitemOpen
  \href@noop {} {}\bibinfo {note} {We are thus assuming that the Hamiltonian is continuously differentiable. This is more than one typically requires for LR bounds, and it does technically rule out circuit models in which the Hamiltonian changes discontinuously from gate to gate, but we view this as a minor restriction since the rate $\zeta$ in Eq.~\eqref{eq:interaction_norm_bounds} can be made arbitrarily large without affecting the dynamical exponent that results.}\BibitemShut {Stop}%
\bibitem [{not()}]{notation_note}%
  \BibitemOpen
  \href@noop {} {}\bibinfo {note} {Throughout the body of the text, we take statements of the form $X \sim Y$ to specifically mean that $\lim X/Y = 1$ in the appropriate limit (e.g., large $h$ in Eq.~\eqref{eq:local_field_distribution}).}\BibitemShut {Stop}%
\bibitem [{ite()}]{iteration_note}%
  \BibitemOpen
  \href@noop {} {}\bibinfo {note} {To start the iteration, $j$ in Eq.~\eqref{eq:strong_field_local_relationship_v2} must be the central site of the right-most obstacle, and we may not have that $j+1 = y$. Thus since the region $\geq j+1$ contains the region $\geq y$, first use that $\lVert \mathcal{P}_{\geq y} A_x(t) \rVert = \lVert \mathcal{P}_{\geq y} \mathcal{P}_{\geq j+1} A_x(t) \rVert \leq 2 \lVert \mathcal{P}_{\geq j+1} A_x(t) \rVert$, i.e., $f_x(y, t) \leq 2 f_x(j+1, t)$, and then begin to use Eq.~\eqref{eq:strong_field_local_relationship_v2}. This explains the prefactor of 2 in Eq.~\eqref{eq:strong_field_repeated_iteration}.}\BibitemShut {Stop}%
\bibitem [{Ale()}]{Alexey_note}%
  \BibitemOpen
  \href@noop {} {}\bibinfo {note} {We would like to thank Alexey V.\ Gorshkov for pointing this out to us.}\BibitemShut {Stop}%
\end{thebibliography}

\end{document}